\documentclass[reprint,aps,nofootinbib]{revtex4-2}

\usepackage[
    top=1.8cm,
    bottom=2cm,
    left=1.5cm,
    right=1.5cm,
    paper=a4paper
    ]{geometry}

\usepackage[utf8]{inputenc}
\usepackage[english]{babel}
\usepackage[T1]{fontenc}
\usepackage[colorlinks=true, breaklinks=true, linkcolor=blue, citecolor=blue, urlcolor=blue]{hyperref}

\usepackage{xcolor}
\usepackage{graphicx}
\usepackage{amsfonts, amsmath, amssymb}

\usepackage{listings}
\newcommand{\be}[1]{\begin{equation}\label{#1}}
\newcommand{\ee}{\end{equation}}

\newcommand{\ket}[1]{| #1 \rangle}
\newcommand{\bra}[1]{\langle #1 |}

\begin{document}

\title{Efficient Emulation of Neutral Atom Quantum Hardware}
\date{\today}

\author{Kemal Bidzhiev}
\email{kemal.bidzhiev@pasqal.com}
\author{Stefano Grava}
\author{Pablo le Henaff}
\author{Mauro Mendizabal}
\author{Elie Merhej}
\author{Anton Quelle}
\email{a.quelle@pasqal.com}

\affiliation{PASQAL, 7 rue Léonard de Vinci, 91300 Massy, France}

\begin{abstract}
Simulating the dynamics of neutral atom arrays is a challenging problem. To address this, we introduce two emulators—emu-sv and emu-mps—as computational backends for Pasqal's pulser package. Emu-sv is designed for high-precision state-vector simulations, giving the possibility to emulate systems of up to $\thicksim27$ qubits on an A100 40GB GPU, making it perfect for cases where numerically exact results are needed. In contrast, emu-mps uses a Matrix Product State representation and other controlled approximations to efficiently simulate much larger arrays of atoms with manageable errors. We show through benchmark comparisons that both emulators provide significant speed-ups over generic solvers such as QuTiP. In addition, we provide practical guidance on choosing between the two emulators. These quantum software tools are designed to support researchers and developers aiming to simulate quantum systems either as a precursor to full hardware implementation or as a means of benchmarking hardware performance.

\end{abstract}

\maketitle

\section{Introduction}
Quantum computing has emerged as a rapidly developing field, offering the potential to outperform classical computers in solving certain classes of problems in optimization~\cite{hogg2000quantum}, material science~\cite{bauer2020quantum}, quantum chemistry~\cite{lanyon2010towards} and machine learning~\cite{biamonte2017quantum}. To build a useful quantum computer, different hardware platforms are being actively developed, including superconducting qubits~\cite{castelvecchi2017quantum, bravyi2022future, gambetta2017building}, trapped ions~\cite{haffner2008quantum, bruzewicz2019trapped}, photonic devices~\cite{o2009photonic, slussarenko2019photonic} and neutral atoms~\cite{henriet2020quantum, lanes2025framework, graham2022multi}. Among these technologies, neutral atom quantum processors have attracted significant attention due to their key properties, including their flexible qubit connectivity,  scalability and long coherence times. In addition, neutral atoms allow for the realization of large arrays of qubits with dynamically reconfigurable geometries and controllable interactions, making them suitable for both quantum simulation and universal quantum computation.

Pasqal's neutral-atom-based QPU~\cite{henriet2020quantum} can be controlled and programmed via the open-source Python package \emph{Pulser}~\cite{silverio2022pulser}, which allows users to build and run neutral atom quantum computing experiments. Pulser enables the creation of the basic building blocks of such experiments, including a virtual register of qubits and laser pulse sequences. Furthermore, Pulser allows for the specification of a customizable noise model to incorporate the various hardware effects, like laser noise or decoherence. To study the resulting quantum dynamics of small systems, Pulser relies on \emph{QuTiP}~\cite{qutip5, qutip2, qutip1} for classical simulations. Although powerful, the QuTiP solver does not scale efficiently to the larger system sizes relevant for neutral-atom processors.

To address this gap, it is crucial to have easy to use \emph{emulators} that faithfully reproduce the features of the Qutip backend, while improving the performance and scalability. In this work, we present two emulators developed for this purpose at Pasqal: \emph{emu-sv}, a state-vector emulator for exact dynamics up to 27 qubits\footnote{Simulations were performed on a NVIDIA A100 GPU 40\,GB memory.}(see Sec.~\ref{sec:hardware}), and \emph{emu-mps}, a Matrix Product State (MPS) tensor network based emulator~\cite{Bidzhiev_pasqal_emulators_2025} designed to handle larger arrays up to a hundred qubits by employing controlled approximations that retain high fidelity. Integrated as numerical solvers for Pulser, they replace the QuTiP backend while remaining fully aligned with the neutral-atom framework.
While internally employing performant numerical techniques, emu-mps and emu-sv are designed for broad usability: they require no specialized knowledge of tensor networks or differential equations and are suitable for users ranging from beginners to expert researchers.
Together, these tools extend Pulser's simulation capabilities from small-scale exact studies to large-scale applications, enabling efficient and accurate analysis of Rydberg-atom dynamics across system sizes of practical interest.

Our emulators have found use in diverse areas of physics and applied mathematics. For instance, emu-mps has been used to compute the dynamical structure factor of systems exceeding $20$ qubits~\cite{vovrosh2025meson}, while both emu-sv and emu-mps have been applied to quantum graph optimization and related problems~\cite{cazals2025quantum, cazals2025identifying}.

The outline of the paper is as follows. First, we detail the dynamics of the quantum system that our emulators solve in Sec.~\ref{sec:hamiltonian-emulator}, and then describe the state vector and MPS representations that are used to encode the associated quantum state in Sec.~\ref{sec:state-rep}. Next, we describe the general features we expect of our emulators in Sec.~\ref{sec:emulators}, and present in details the emulators: emu-sv in Sec.~\ref{sec:emusv} and emu-mps in Sec.~\ref{sec:emumps}. This detailed description includes performance and memory benchmarks. Finally, we conclude in Sec.~\ref{sec:discussion}.

\section{Numerical simulation of Quantum Dynamics}\label{sec:numerical-quantum}

\subsection{Time-evolution of a Quantum System}\label{sec:hamiltonian-emulator}

While the atoms in Pasqal's hardware naturally possess rich multiple energy levels, we focus here on a simplified and effective two-level description for each atom, modeling the system as an array of \emph{qubits}. This approach retains the essential computational features while simplifying the simulation framework. The Hilbert space of states with $N$ qubits then reads
\be{eq:statespace}
\mathcal{H} = \otimes_{i=0}^N \mathbb{C}^2 \simeq \mathbb{C}^{2N},
\ee
where $\mathbb{C}^2$ denotes the Hilbert space of a single qubit, defined with respect to the computational basis $\{\vert0\rangle, \vert1\rangle\}$. This basis is used throughout the paper unless stated otherwise.

The time evolution of a quantum state $\ket{\psi} \in \mathcal{H}$ is governed by the time-dependent \emph{Schr\"odinger equation} (expressed in natural units with $\hbar = 1$):
\be{eq:schro}
i\frac{d \ket \psi }{d t} = \hat H(t) | \psi \rangle,
\ee
where we consider a particular time-dependent \emph{Hamiltonian} operator 
\be{eq:hamiltonian}
    \hat{H}(t) =  \sum_{i=1}^N \frac{\Omega_i(t)}{2} \hat \sigma^x_i -  \sum_{i=1}^N \delta_i(t) \hat n_i + \sum_{i>j} \frac{C}{|\mathbf r_{ij}|^6} \hat n_i \hat n_{j}.
\ee
Here, $\hat n=\ket{1}\bra{1}$ is the orthogonal projection onto $\ket{1}$ and $\hat \sigma^x =\ket{1}\bra{0}+\ket{0}\bra{1}$ is the Pauli $\text{X}$ matrix. The subscript $i$ indicates that the operator nontrivially acts on the $i$-th qubit, while acting as the identity on the other qubits. The third term describes the pairwise van der Waals interaction between qubits $i$ and $j$, with $\mathbf{r_{ij}}$ denoting distance between qubits and $C$ is a constant depending on the hardware specification. The control parameters $\Omega_i(t)$ and $\delta_i(t)$ are the Rabi frequencies and detunings, respectively, and make the dynamics programmable.

\subsection{Representing quantum states}\label{sec:state-rep}

Pasqal's emulators make use of two different representations of quantum states. A quantum state \( \ket{\psi} \) is fully determined by its amplitudes:
\begin{equation}
    \ket{\psi(t)} = \sum_{i_1 \dots i_N} c_{i_1 \dots i_N}(t) \ket{i_1 \dots i_N},
    \label{eq:wf_state}
\end{equation}
which can be represented either as an $N$-dimensional tensor with components $ c_{i_1 \dots i_N} $, or, equivalently, as a vector of length $2^N$, due to the isomorphism described in Eq.~\eqref{eq:statespace}. Since both approaches share the same memory layout in CPUs and GPUs of a classical computer, e.g. in CUDA~\cite{sanders2010cuda} and in derived packages such as Pytorch~\cite{paszke2019pytorch, imambi2021pytorch}, which the emulators rely upon, we refer to both as the state-vector representation.

\subsubsection{State-vector representation}
The state-vector representation offers the advantage of solving Eq.~\eqref{eq:schro} directly using various established numerical methods. This makes state-vector emulators both reliable and faster than MPS-based emulators for systems with small qubit numbers. However, this approach faces a fundamental limitation, as the dimension of the Hilbert space $\mathcal{H}$ scales exponentially with $N$, restricting the maximum number of qubits that can be simulated in principle.

To circumvent this limitation, alternative state representations have been developed, among which we discuss the Matrix Product State representation.

\subsubsection{MPS representation}
 The tensor of amplitudes $c_{i_1\dots i_N}$ in Eq.~\eqref{eq:wf_state} factorizes using the MPS representation~\cite{Schollwck2011, Oseledets2011} as follows:
\be{eq:mps}
c_{i_1 \dots i_N} = A^{i_1}_{j_1} A^{i_2}_{j_1,j_2} \dots A^{i_N}_{j_N},
\ee
where the Einstein summation convention is used~\cite{orus2014tn, perez2006matrix}. The indices $j_k$ are called \emph{bond indices}, and the range of a given bond is called the \emph{bond dimension}. Such a factorization always exists and it is highly non-unique. The memory requirement of an MPS is determined by the bond dimensions. Highly entangled states might require more memory in MPS representation compare to the full state vector description. However, the states typically encountered in the emulation problems outlined above are low entangled states and can usually be expressed with considerable memory savings. For example, product states, like $\ket{0\dots 0}$, which is used as an initial state for simulating the neutral atom QPU, have an MPS representation with all bond dimensions equal to $1$, resulting in an exponential reduction in memory consumption as a function of $N$. In return for these memory savings, special care is required when solving Eq.~\eqref{eq:schro} to avoid creating intermediary MPS states with excessively large bond dimension. We discuss this issue further in Sec.~\ref{sec:emumps}, where we describe the specific algorithms used to address it.

\section{The emulators}\label{sec:emulators}
Having introduced the two representations used by Pasqal's emulators, we now turn to the corresponding emulators: emu-sv (\emph{emu}lation - \emph{s}tate \emph{v}ector) and emu-mps (\emph{emu}lation - MPS). Although the two emulators differ in their internal algorithms and performance characteristics, they share many common features, as both are designed to emulate a QPU based on neutral atoms. We therefore begin with a discussion of the features common to both emu-sv and emu-mps.

Predominantly, integration of the emulators into the existing software ecosystem of Pasqal is a concern. In particular, they must be compatible with Pulser~\cite{silverio2022pulser} and with existing machine learning Pytorch~\cite{paszke2019pytorch} workflows. Additionally, it should be easy to switch between these emulators in a workflow, or even to a remote version of one of these emulators run in the cloud~\cite{emutn}. To conveniently meet these requirements, both emulators implement the API defined by Pulser and use PyTorch as the backend for numerical computations. This design choice offers a uniform usage of the emulators, with differences arising only in emulator-specific configuration options. For instance, many MPS-based algorithms rely on truncation techniques, and thus emu-mps requires a truncation tolerance parameter, whereas a emu-sv does not.

In addition, both emulators share the same computational model. Pulser is primarily a software package for defining the time-dependent control parameters $\Omega_i(t)$ and $ \delta_i(t)$ in the Hamiltonian. These parameters are specified as functions of time using a small set of basic building blocks, including linear functions, the Blackman window function~\cite{podder2014comparative}, and cubic splines that interpolate a given set of points ~\cite{mckinley1998cubic, knott1999interpolating}. Pulser also offers functionality for sampling these functions, giving values such as $\Omega_{i,n} = \Omega_i(t_n)$ at discrete times $t_n= n$ ns, and similarily for $\delta_{i,n}=\delta_{i}(t_n)$.

For computational purposes, both emulators approximate the given continuous and smooth pulse with a piecewise constant one, obtained using the sampling mechanism mentioned above. Each emulator takes a configuration parameter $dt$, corresponding to an integer multiple of the $1\text{ns}$ sampling resolution, and assumes the pulse to be piecewise constant for this duration. For maximum order of the resulting integration procedure, parameter values are taken at the midpoint of each time step: let $m = (n + \frac{1}{2}) dt$ then for time step $n$ we use the parameter values $(\Omega_{i,\lfloor m \rfloor} + \Omega_{i, \lceil m \rceil})/2$, and analogously for $\delta_i$. The pulse discretization allows to approximate the time-dependent Hamiltonian $H(t)$ in Eq.~\eqref{eq:schro} with a set of constant Hamiltonians $H_k$, where solutions for each time step can be obtained as
\be{eq:timestep}
\ket{\psi(t_{k+1})} = \ket{\psi(t_k+dt)} = e^{- i H_k dt} \ket {\psi (t_k)}.
\ee
Thus, the task of the emulator reduces to efficiently computing the right-hand side (RHS) of Eq.~\eqref{eq:timestep}. Such a discretization makes the integrator formally second order in $dt$, which is not especially high. However, higher-order errors arise only from the time-dependent nature of the Hamiltonian. As we will demonstrate in later sections, such a scheme yields acceptable emulator accuracy even for $dt > 1 \rm{ns}$. For this reason, we have decided against increasing the integration order through methods such as Richardson extrapolation~\cite{hairer2010solving}.

\subsection{emu-sv}\label{sec:emusv}
When using the state-vector representation to compute the RHS of Eq.~\eqref{eq:timestep}, one quickly finds how expensive it is to store the Hamiltonian in dense format~\cite{H_ner_2017}. Fortunately, the Rydberg Hamiltonian in Eq.~\eqref{eq:hamiltonian} possesses considerable structure that can be exploited to outperform existing sparse matrix formats such as the COO and CSR formats~\cite{chou2018format}. Specifically, the only non-diagonal entries in Eq.~\eqref{eq:hamiltonian} are generated by the $\sigma^x_i$ terms, which act by swapping subvectors of the state-vector. As a result, it suffices to store the values $\Omega_{i,n}$ and only the diagonal of $H$. Such a representation requires less memory than both the COO and CSR formats.

The matrix-vector multiplication, i.e. a Hamiltonian $H_k$ application to a state $\ket{\psi(t_k)}$, is then defined in two steps: 
first, by applying the diagonal of $H_k$, and 
second, by separately adding the appropriate permutations of $\ket \psi$ for $\sigma^x_i$, scaled by $\Omega_{i,n}$. This implementation is slightly faster than \texttt{torch.matmul} with the CSR format, which itself vastly outperforms the COO format for large matrices. With matrix-vector multiplication available, one could in principle use any general ODE solver to solve Eq.~\eqref{eq:schro} directly.
However, given that computing the RHS of Eq.~\eqref{eq:timestep} provides a good approximation to the ideal solution even for larger $dt$, it becomes preferable to use the Lanczos algorithm~\cite{komzsik2003lanczos, hochbruck1997krylov, noack2005diagonalization}, which is significantly faster than Runge-Kutta or multistep methods at large $dt$~\cite{cartwright1992dynamics}, as it can reach very high integration orders.
The trade-off, however, is memory: in the Lanczos algorithm, each additional order requires storing one more Krylov vector, which is of size $2^N$, so this choice exchanges memory efficiency for speed.

In addition to simulating clean, noiseless systems, emu-sv can also model open-system dynamics using the same noise mechanisms implemented in Pulser. 
Depending on the specific model, this is done either by varying Hamiltonian parameters on a shot-to-shot basis or by solving the full Lindblad master equation, with time evolution performed using the Lanczos algorithm.

\subsubsection{Performance}

We compare the performance of emu-sv, in terms of both runtime and accuracy, against the QuTiP-based emulator implemented in Pulser~\cite{qutip5, qutip1, qutip2}. In the time-dependent case, QuTiP solves Eq.~\eqref{eq:schro} using the ZVODE ODE solver~\cite{VODE, hairer2010solving}, which is an multistep method formulated in Nordsieck form. This allows the solver to adapt both the step size and the integration order, in contrast to the fixed-step approach of emu-sv.
The input for the solver ZVODE is obtained by taking the discretized Hamiltonian parameters $\Omega_{i,n}$ and $\delta_{i,n}$, which are then interpolated cubically to provide a smooth input. While the integration order of ZVODE is adaptive but capped at 12, and as the number of qubits increases, the spectral width of the Hamiltonian grows, requiring increasingly finer time steps $dt$.
By contrast, the Lanczos algorithm employed in emu-sv effectively achieves an adaptive order by building a Krylov subspace until the algorithm is converged, thereby avoiding the fixed upper limit of ZVODE. As a result, even for large qubit systems with broad Hamiltonian spectra, the method can advance the dynamics with comparatively larger time steps, leading to improved performance
\begin{figure}[ht!]
    \centering
    \includegraphics[width=1.0\linewidth]{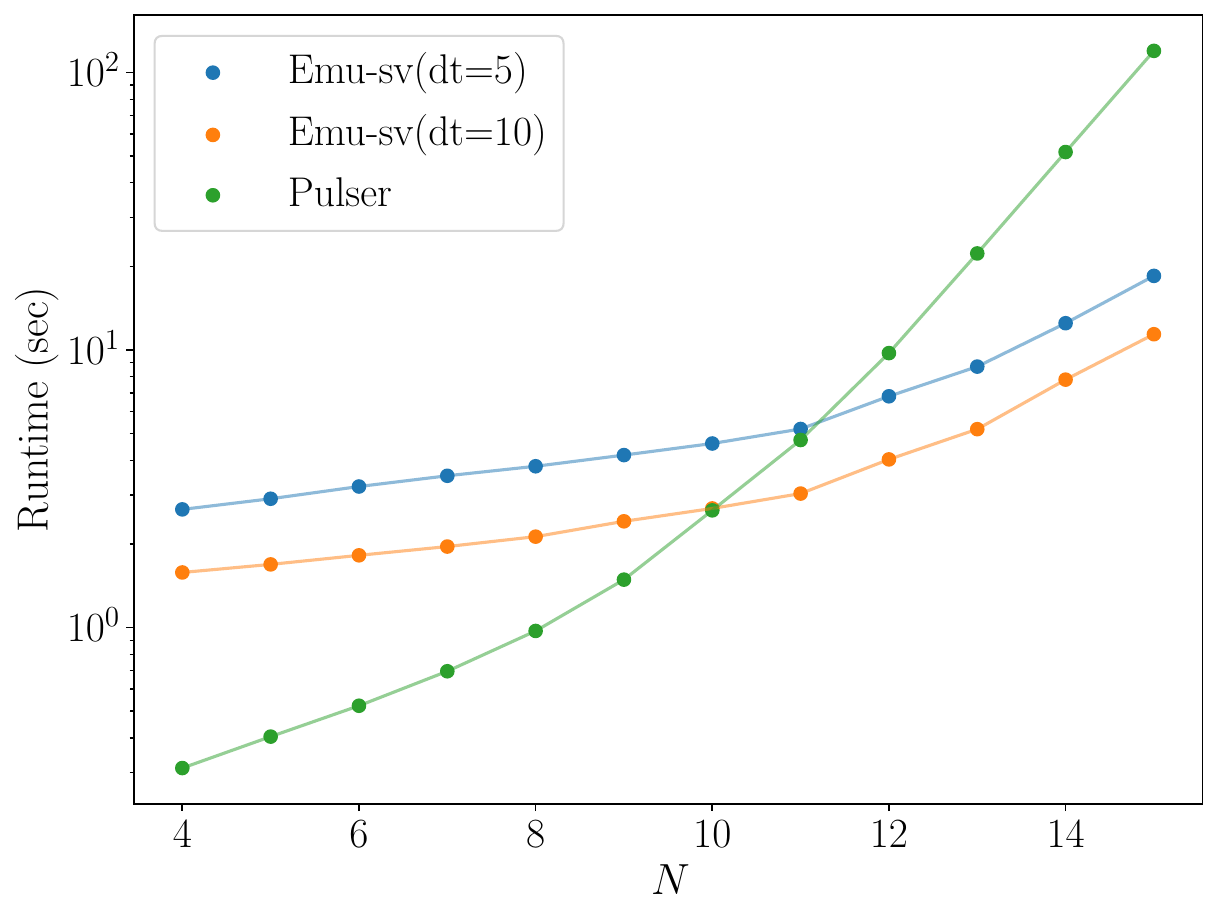}
    \caption{Comparison of runtimes for different system sizes $N$ of emu-sv with $dt=5, 10$ and Pulser. Pulsers default backend QuTiP uses ZVODE ODE solver. From about 9 qubits onwards, runtime approximately doubles for each extra qubit for Pulser, which is to be expected matrix-vector multiplication starts dominating the runtime of the program. Emu-sv uses pytorch native parallelization tools with number of threads = 16. The same exponential scaling as for QuTiP sets in for emu-sv (see Fig.~\ref{fig:sv-gpu}), but it does so later because the solver is more efficient for larger matrix sizes.}
    \label{fig:sv-pulser}
\end{figure}

\begin{figure}[ht!]
    \centering
    \includegraphics[width=1.0\linewidth]{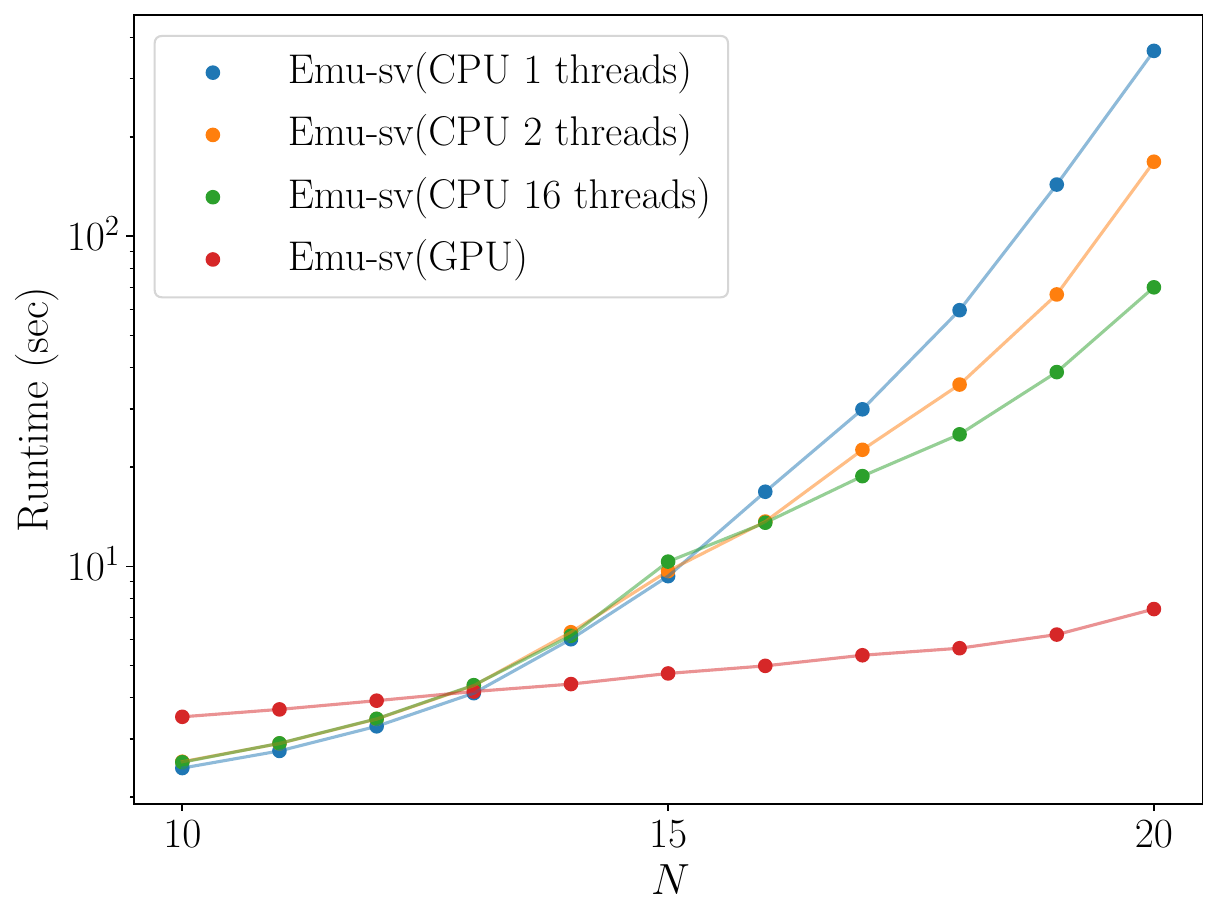}
    \caption{Comparison of runtimes for emu-sv when running on GPU and CPU with 1, 2, and 16 threads. Due to high parallelism, larger system instances are executed faster on GPUs. However, the situation is the opposite for smaller system sizes, the individual cores in a GPU are slower, and the parallelism cannot be fully leveraged.}
    \label{fig:sv-gpu}
\end{figure}

Memory consumption can be efficiently estimated by considering the integration order of the Lanczos algorithm required to perform a time step. As mentioned above, emu-sv stores the current state $c_{i_1\dots i_N}(t_k)$ and the diagonal of the Hamiltonian $H(t_k)$, each of size $2^N$. The peak memory consumption in emu-sv occurs during the Lanczos procedure, in which all Krylov vectors up to the required order are stored to achieve convergence. Each Krylov vector has length $2^N$. Since emu-sv uses complex double precision numbers (64-bit arithmetic) a state vector requires $16 \times 2^{N}$ bytes of memory, corresponding to approximately $1$ gigabyte (GB) for $26$ qubits. Assuming $15$ Krylov vectors are required for convergence, the total memory consumption using emu-sv remains below $20$ GB for $26$ qubits.

\begin{figure}[ht!]
    \centering
    \includegraphics[width=1.0\linewidth]{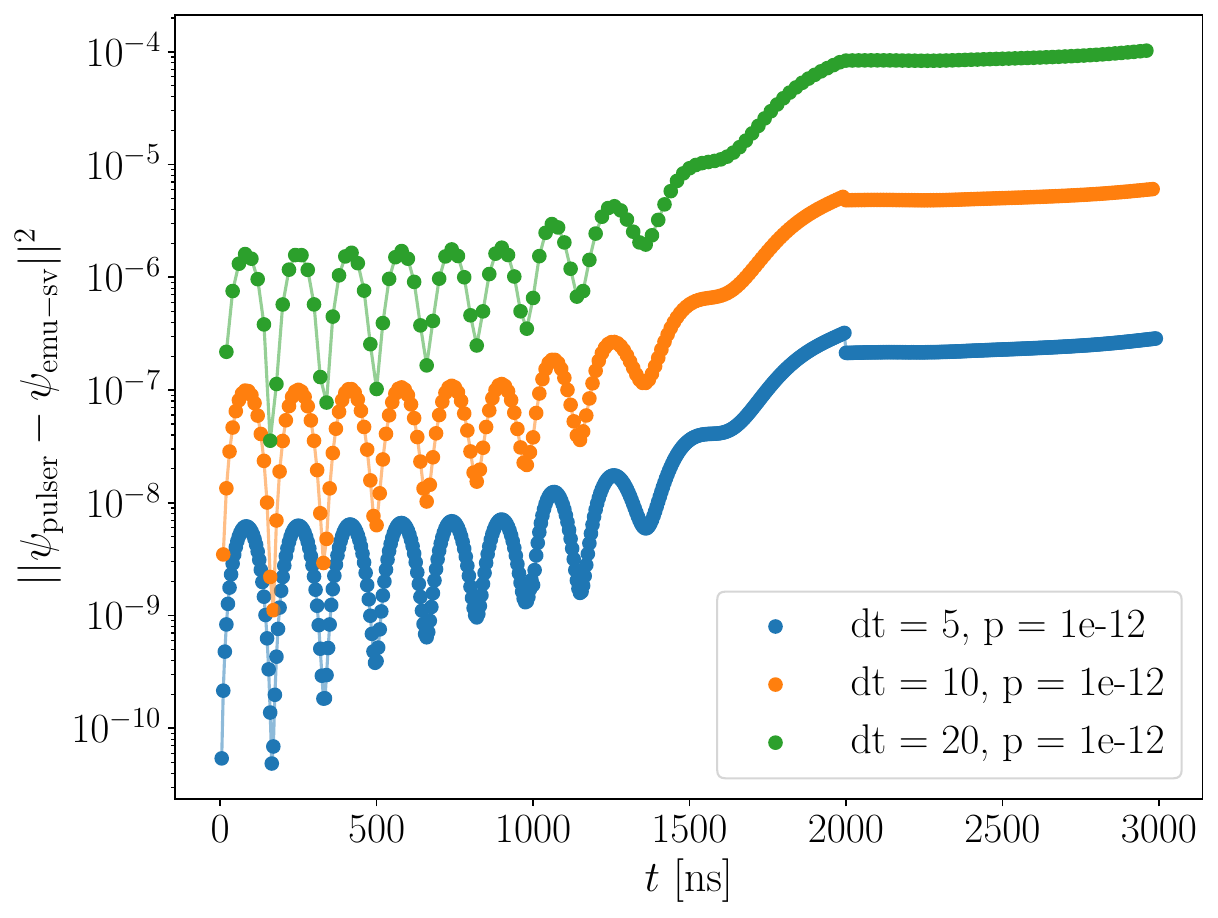}
    \caption{Time evolution of the norm difference between the pulser and emu-sv wavefunctions. The precision parameter is fixed at $p = 10^{-12}$ to highlight that the error is governed predominantly by the discretization scheme, determined solely by the time step $dt$. Each application of the time evolution with a given $dt$ introduces an error that accumulates over time.}
    \label{fig:sv-fidelity-time}
\end{figure}

\begin{figure}[ht!]
    \centering
    \includegraphics[width=1.0\linewidth]{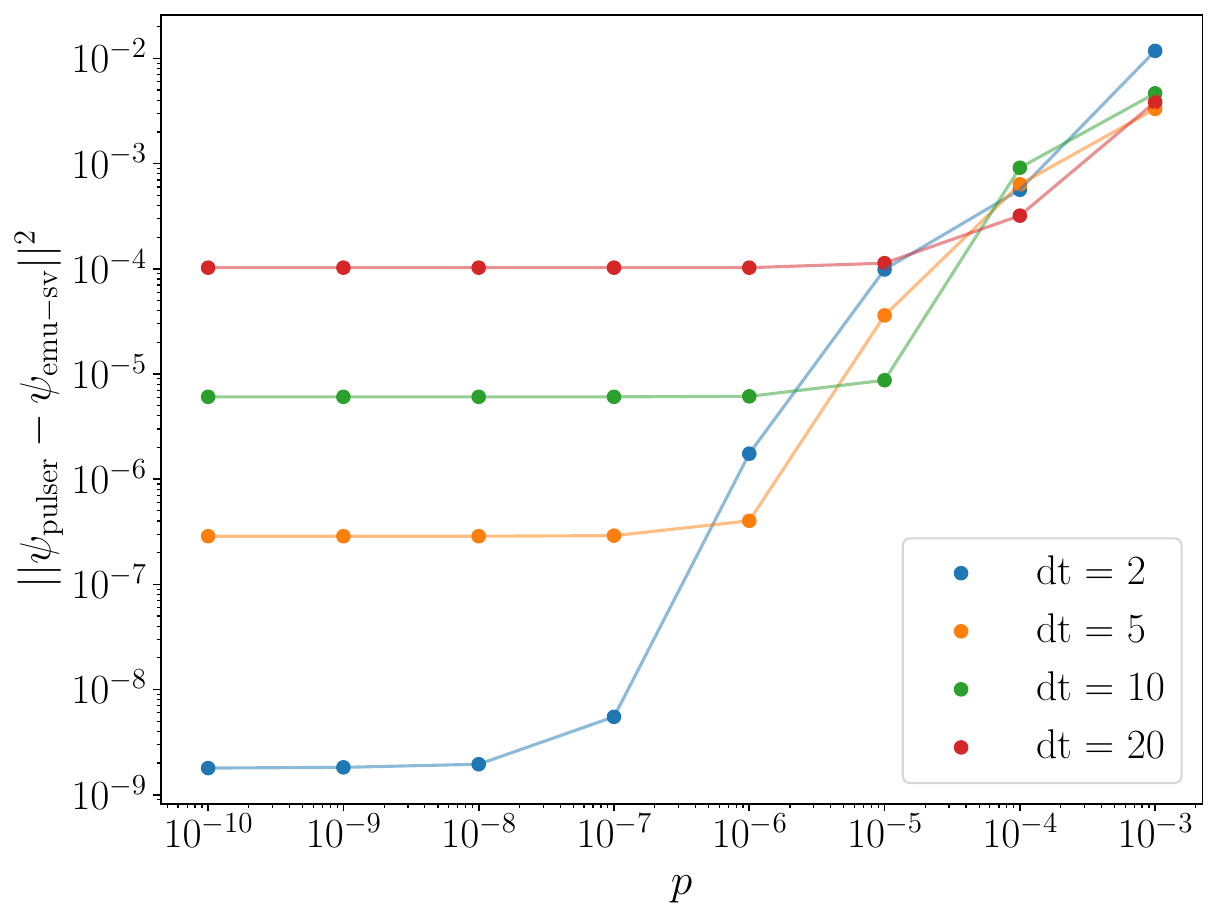}
    \caption{Norm of the difference between pulser and emu-sv wavefunctions at the end of the time evolution as a function of the precision parameter $p$, evaluated over a logarithmic range from $10^{-10}$ to $10^{-3}$. Each different curve corresponds to time step size $dt$, illustrating the dependence of the error on both the precision parameter and the time discretization.
    }
    \label{fig:sv-fidelity-final}
\end{figure}

Finally, we discuss accuracy. As mentioned above, our solver operates at relatively low order. However, this is acceptable for typical use cases at Pasqal. To make this claim more precise, in Figs.~\ref{fig:sv-fidelity-time} and ~\ref{fig:sv-fidelity-final} we benchmark emu-sv against Pulser using a 9-qubit adiabatic sequence that reflects quantum annealing problems commonly simulated at Pasqal.

\subsection{emu-mps}\label{sec:emumps}
For emu-mps, the quantum state is represented as a Matrix Product State, while the Hamiltonian operator is represented as a Matrix Product Operator (MPO). The right-hand side of Eq.~\eqref{eq:timestep} is then computed using the Time-Dependent Variational Principle (TDVP)~\cite{Haegeman2011, haegeman2016unifying}. Specifically, emu-mps employs a second-order two-site TDVP scheme. A central part of this algorithm is the sequential time evolution of two-site subsystems using a reduced effective Hamiltonian. This procedure is equivalent to applying a matrix exponential to a vector, therefore, the same Lanczos routine is employed. 
Additionally, emu-mps supports the same noise models as Pulser\cite{emulators_2025}, but rather than solving the full Lindblad master equation as in emu-sv, it implements noise via Monte Carlo quantum jumps~\cite{molmer1993monte}, thereby using stochastic sampling (quantum trajectories) on pure matrix product states to approximate open-system dynamics.

\subsubsection{Performance}

\begin{figure}[ht!]
    \centering
    \includegraphics[width=1.0\linewidth]{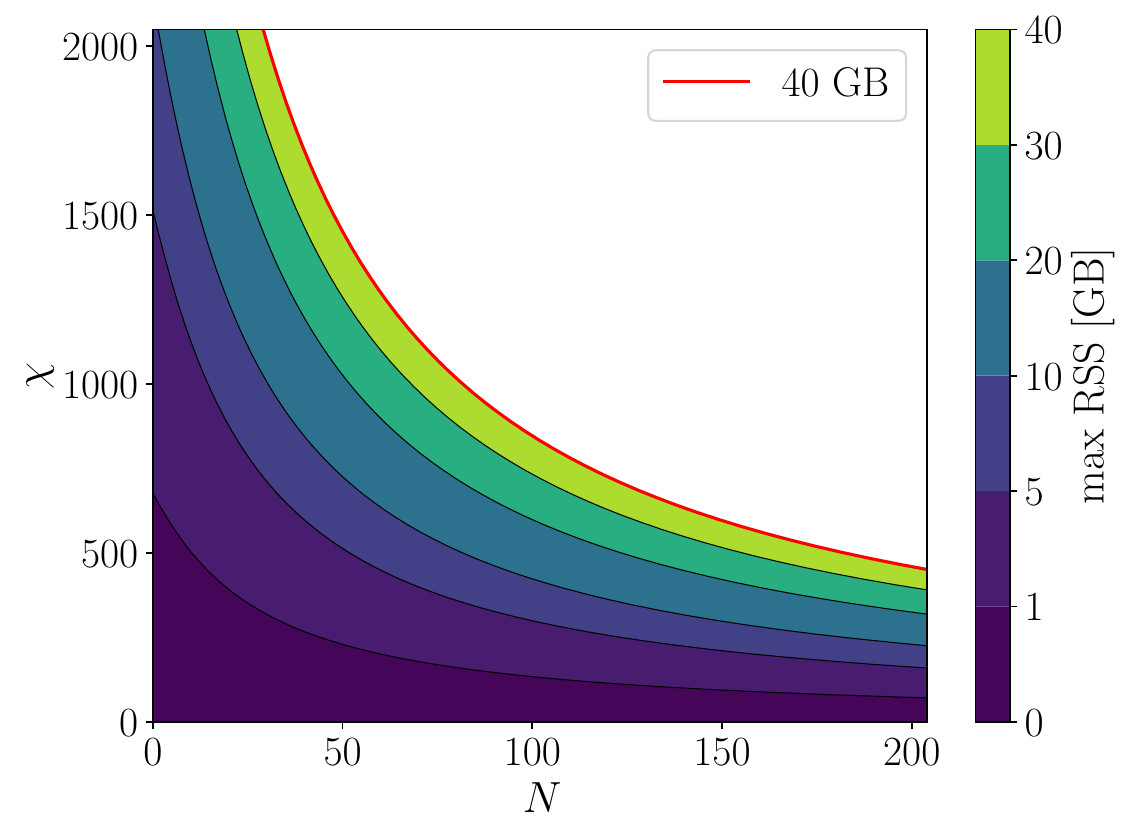}
    \caption{Upper bound on the memory used by emu-mps to simulate a given number of qubits $N$ with a given maximum bond dimension $\chi$~\cite{emulators_2025}. The number of Lanczos iterations required for convergence is set to $k = 30$ for this plot. This bound is slightly pessimistic, and actual values depend on the true number of iterations, which are influenced by $dt$ and the tolerance for convergence of the Lanczos algorithm.}
    \label{fig:mps-memory}
\end{figure}

\begin{figure}[ht!]
    \centering
    \includegraphics[width=1.0\linewidth]{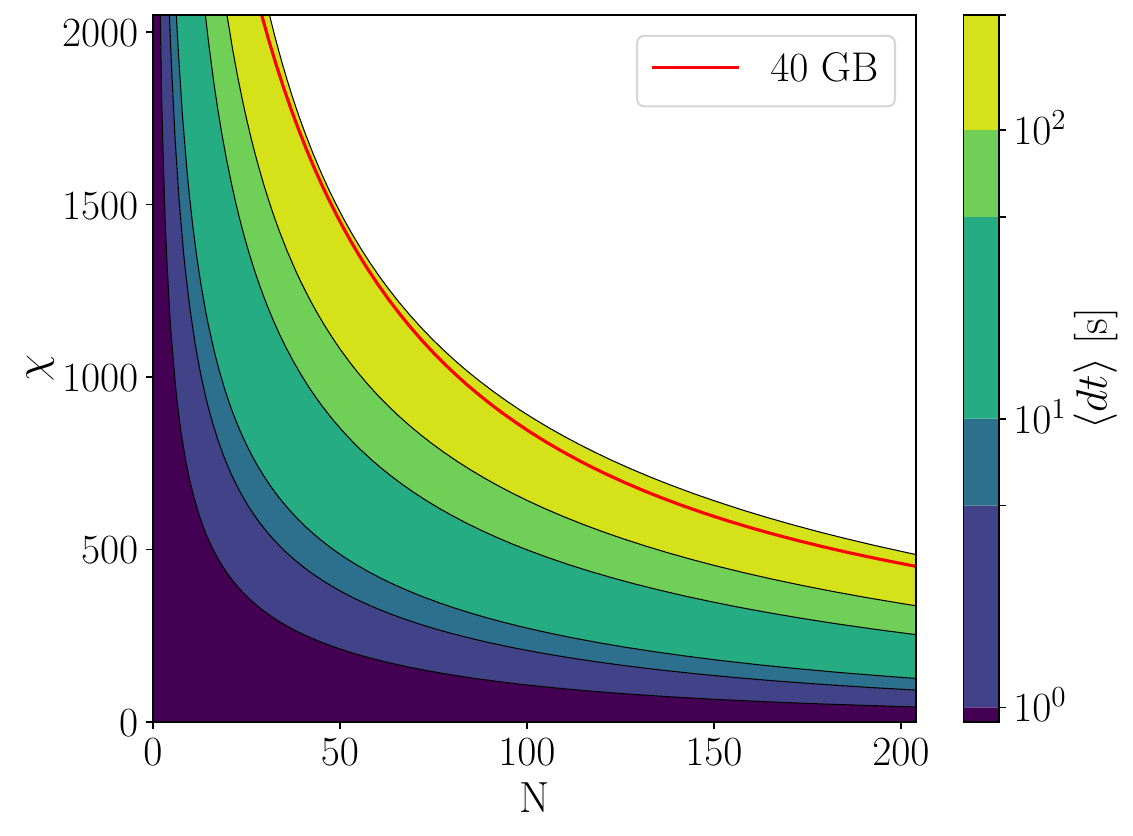}
    \caption{Estimation of the time needed by emu-mps to perform a single timestep for a given number of qubits $N$ and a given maximum bond dimension $\chi$. In principle, the number of iterations required by the Lanczos algorithm to converge is a hidden dependent variable. We've found it to depend only weakly on $N$ and $\chi$ for the plotted range of parameters, see also our estimate of 30 in Fig.~\ref{fig:mps-memory}.}
    \label{fig:mps-runtime}
\end{figure}

\begin{figure}[ht!]
    \centering
    \includegraphics[width=1.0\linewidth]{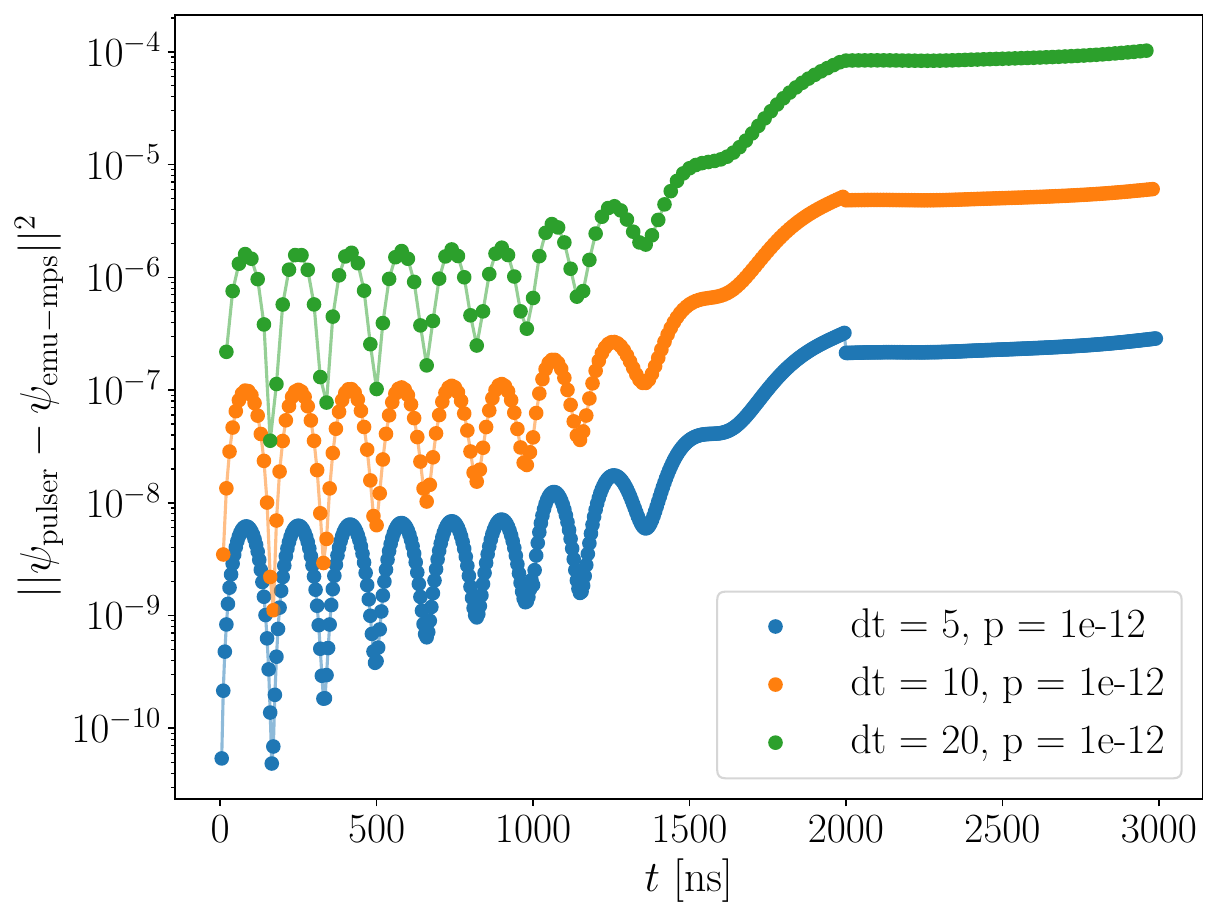}
    \caption{Time evolution of the norm difference between the pulser and emu-mps wavefunctions. The MPS truncation parameters $\chi = 1000$ and $p = 10^{-12}$ are fixed to emphasize the discretization scheme error as in Fig.~\ref{fig:sv-fidelity-time}. In a similar way, the error accumulates with time and grows by order of magnitude by the end of the evolution.}
    \label{fig:mps-fidelity-time}
\end{figure}

Memory usage of emu-mps is divided among three main parts: the time-evolved MPS, the bath tensors required by TDVP, and intermediate tensors representing the basis for the Krylov subspace generated in the Lanczos algorithm. Therefore, the peak memory consumption can be estimated from the configuration parameters. The details of this calculation are described in detail in~\cite{emulators_2025}. This implies that one can predict in advance whether a given simulation will fit in memory based on the number of qubits and the maximum bond dimension of the MPS. The result is illustrated graphically in Fig.~\ref{fig:mps-memory}.
Similar considerations apply to estimating the runtime. As with memory estimation, the details are provided in the documentation \cite{emulators_2025}, and the result are shown in Fig.~\ref{fig:mps-runtime}.
Regarding the accuracy of the emulator, we discussed in Sec.~\ref{sec:emusv} the discretization errors, which were found to be on the order of $10^{-5}$ for a typical adiabatic sequence. In Fig.~\ref{fig:mps-fidelity-time} we show the analogous result for emu-mps. 
\begin{figure}[ht!]
    \centering
    \includegraphics[width=1.0\linewidth]{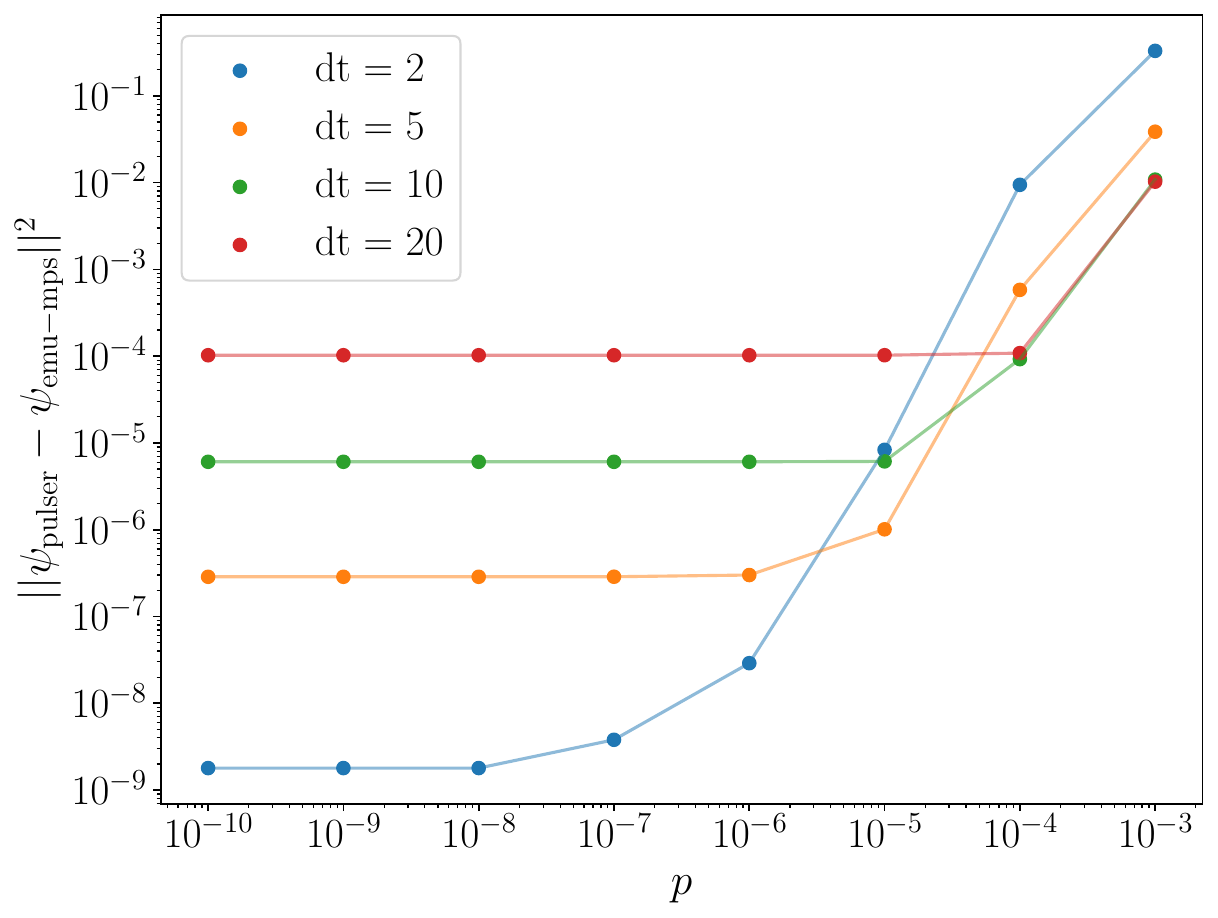}
    \caption{Norm of the difference between pulser and emu-mps wavefunctions at the end of the time evolution. 
    The truncation parameter $p$ varies over a logarithmic range from $10^{-10}$ to $10^{-3}$, while the maximal allowed bond dimension is fixed at $\chi = 1000$.  For any $dt$, increasing $p$ results in the truncation error from small singular values becoming the dominant contribution, surpassing the discretization error.}
    \label{fig:mps-fidelity-final}
\end{figure}

The default truncation precision $p$ for the MPS is $10^{-5}$. In the second-order TDVP scheme, $2N-3$ truncations are applied at each timestep, so after $m$ steps the total error can be bounded by $p \times m \times (2N-3)$. However, subsequent truncation errors are likely to partially cancel, and the observed total error is typically much smaller, see Fig.~\ref{fig:mps-fidelity-time} and~\ref{fig:mps-fidelity-final}. When the simulation reaches the maximum bond dimension allowed for the MPS, control over the truncation error is lost in order to satisfy this bounds, and the accuracy could become different. In such cases, care must be taken to ensure the reliability of the results.
 
This discussion ignores the presence of errors inherent in the TDVP algorithm~\cite{haegeman2016unifying, kloss2018time}. When simulating two-dimensional systems, a potentially large source of errors in TDVP arises from the effective description of qubits that are not neighboring in the ordering of MPS sites. For instance, on square lattices, qubits that are physically close are unavoidably distant from the MPS perspective, which introduces additional errors in the emulator. 
We have found that choosing a good \emph{qubit ordering} in the MPS is crucial for the qualityof the results\cite{Dauphin2024, qubits_ordering}, even for system sizes where emu-sv can also simulate, and where truncation of the MPS is not needed. Since finding an optimal ordering is generally non-trivial, or in some cases impossible, emu-mps can be configured to automatically attempt an improved qubit ordering using the Cuthill-McKee algorithm~\cite{cuthill1969reducing, Dauphin2024}. It is important to note that such a reordering can also reduce the bond dimension required to capture the essential physics.

\section{Discussion}\label{sec:discussion}
As can be seen from the discussion in Sec.~\ref{sec:emulators}, emu-sv and emu-mps have different performance characteristics as a result of the different state representation used. 
It can be seen that emu-mps abstracts away much of the details of the underlying MPS representation~\cite{emulators_2025}, at least when it comes to common workflows encountered within Pasqal. As a result, the usage of emu-sv and emu-mps is highly uniform, and it is possible to switch between the two emulators in a workflow with minor impact on the code to leverage the different performance characteristics for optimal results. Based on our benchmarks, which were performed in the environment outlined in appendix~\ref{sec:hardware}, we can offer the following rules of thumb when running on GPU:

\begin{center}
\begin{tabular}{ |c|c| } 
\hline
qubit number &  emulator \\
\hline
$\leq 27$ & emu-sv \\ 
$> 27$ & emu-mps \\ 
\hline
\end{tabular}
\end{center}

Intuitively, as mentioned in Sec.~\ref{sec:state-rep}, emu-sv is faster than emu-mps because its algorithms are less involved, but due to exponential scaling in memory and runtime, it will not be able to go beyond approximately $27$ qubits (depending on available memory). Finally, since emu-mps makes various approximations in its core algorithms, it must be configured correctly, and is in principle less accurate, so comparing results between the emulators for $\approx 25$ qubits is valuable.

Regarding possible future directions of development for the emulators, we are mostly guided by the use-cases identified within the company.  For example, we already mentioned building the package upon torch to satisfy machine learning requirements. Pasqal currently has not open sourced any machine learning packages built upon the emulators described here, and for this reason, we have omitted a detailed discussion of differentiability. However, emu-sv is differentiable, and plans exist to make emu-mps differentiable also. What this means, is that when one defines the pulser sequence to simulate in terms of differentiable torch tensors~\cite{paszke2019pytorch}, then one can compute the derivative of emulator output such as expectation values w.r.t these tensors, and use these gradients in machine learning workflows. As another example, it is of interest to follow the ground state of Eq.~\eqref{eq:hamiltonian} over time, so we are implementing the Density Matrix Renormalization Group (DMRG)~\cite{schollwock2011density, verstraete2023density} algorithm in emu-mps to track this ground state over time. Finally, the teams working on the hardware are continually analyzing the sources of noise in our devices, and when new sources become relevant, for example because a dominant noise source is reduced in magnitude, they will have to be added to both Pulser and the emulators described here.

\begin{acknowledgements}
We thank Louis-Paul Henry and Joseph Vovrosh for their suggestions regarding the manuscript, and the emulator users for their feedback and help improving the product.
\end{acknowledgements}

\appendix

\section{A note on the cluster hardware}\label{sec:hardware}
The benchmarks mentioned in this paper were run on either a AMD EPYC 7742 CPU, or the 40GB DRAM version of the A100 NVidia GPU. When run on the cpu, up to 1TB of RAM is available, but this has to be shared with other users of the cluster node.
The simulations using pulser were based on version 1.5 backed by qutip 5.2, numpy 2.2 and scipy 1.15.
The simulations using our emulators were done on version 2.3 backed by torch 2.8.


\bibliographystyle{unsrtnat}
\bibliography{refs} 

\end{document}